# Low Temperature Spectral Dynamics of Single Molecules in Ultrathin Polymer Films

Yaroslav I. Sobolev, [1*] Andrei V. Naumov, [1,2] Yuri G. Vainer, [1,3] Lothar Kador, [4]

[1] *Institute for Spectroscopy, RAS, Troitsk, Moscow region, 142190, Russia*
[2] *Moscow Pedagogical State University, Malaya Pirogovskaya str. 29, 119992 Moscow, Russia*
[3] *Moscow Institute of Physics and Technology, Dolgoprudny, Russia*
[4] *Institute of Physics and BIMF, University of Bayreuth, 95440 Bayreuth, Germany*

We studied the spectral dynamics of single fluorescent dye molecules embedded in ultrathin films (5 – 100 nm) of the amorphous polymer polyisobutylene at cryogenic temperatures and its variation with film thickness. Noticeable portion of molecules in the ensemble show a behavior which is inconsistent with the standard tunneling model: Their spectral lines are subject to irreversible spectral jumps, continuous shifting, and abrupt chaotic changes of the linewidth or jumping rate. In films thinner than 100 nm, the occurrence of "non-standard" spectral behavior increases with decreasing sample thickness at fixed excitation intensity. In addition, it also increases with laser intensity.

The question as to how and to which degree the properties of a material are influenced by the size and shape of an object made of it is of fundamental importance in many fields of physics. It has been attracting more and more attention in recent years, since the possibilities of creating and studying nano-scale objects are growing at a fast rate.

According to many experiments, properties of very thin polymer films differ in many ways from those of the corresponding bulk materials. For example, in ultrathin polystyrene films the viscosity is higher by several orders of magnitude and has a smaller temperature dependence,[1] relaxation processes are faster on the free surface and slower near the supporting substrate,[2] and the density of the vibrational modes constituting the boson peak is lower.[3] Furthermore, these size effects are different for free-standing films and films deposited on a rigid substrate. It appears that in the latter case the local behavior of the polymer near the substrate is distinctly different from that in the vicinity of the free surface. For example, the mobility of the polymer chains is lower close to the substrate ("dead layer") and increases towards the free surface.[3,4] The influence of the two bounds – substrate and free surface – extends several ten nanometers into the material. Therefore, the properties of a supported film of several ten nanometers thickness are strongly inhomogeneous and feature a smooth transition from those at the free surface to those near the substrate. It is not readily obvious which influence prevails for a given film thickness.

The complexity of thickness effects in polymers is best illustrated by the investigations of the glass transition temperature ($T_g$) as a function of film thickness. As noted by McKenna,[5] there would probably be little general interest in the influence of film thickness on $T_g$, if $T_g$ increased in thin samples. But experiments have revealed both a modest decrease of $T_g$ in ultrathin films supported on rigid substrates[6] and very large reductions (more than 60 K) in free-standing films of polystyrene.[7] At present, no generally accepted theory of the glass transition dynamics can quantitatively explain these findings. Even the experimental results themselves are not unambiguous. Different methods of measuring $T_g$ in thin films sometimes yield drastically different results (see Table 2 in Ref.[8]). Furthermore, it was shown that thorough annealing removes any dependence of $T_g$ on film thickness or molecular weight.[8] This implies that the thickness effects depend on sample history as well, making the problem even more complex.

This unsettled experimental situation of polymers in confined geometry, along with the scarcity of low-temperature investigations,[2,9,10] was our motivation to study the impurity spectral dynamics in thin polymer films at cryogenic temperatures of a few K. This should provide a new approach to the problem, since the dynamics is significantly simplified at low temperatures. In order to demonstrate the differences of the behavior in thin films as compared to that in the bulk, we will briefly outline the latter, i.e., the usual low-temperature dynamics of polymers.

The dynamics of amorphous solids (organic and inorganic glasses, polymers, resins, etc.) at temperatures of a few K is usually described within the framework of the standard tunneling model of glasses. This universally accepted phenomenological model is based on the concept of independent ''tunneling two-level systems'' (TLSs) – elementary localized non-interacting low-energy excitations intrinsic to all disordered solids.[11,12] Each TLS corresponds to tunneling transitions of an atom (or a group of atoms) between two localized potential minima. According to this model, the density of TLSs far exceeds the density of acoustic phonons at T < 1 K, so the TLS contribution to the overall dynamics prevails. At temperatures between a few K and a few ten K, the dynamical phenomena in amorphous materials become more intricate and no longer comply with the standard tunneling model of glasses. Instead, they can be described with the concept of quasi-localized low-frequency vibrational modes (LFMs).[13,14] These excitations correspond to vibrations which are localized (i.e., non-propagating) due to the high degree of local inhomogeneity in disordered solids.

Until lately, the low-temperature dynamics of amorphous solids seemed to be surprisingly universal: it appeared to be qualitatively independent of the structure and chemical composition of the material and was believed to arise from local disorder and heterogeneity itself. Nonetheless, more and more experimental results are being published that cannot be fully explained within the framework of the standard tunneling model of glasses.

The spectroscopy of single impurity centers in solids (single-molecule spectroscopy, SMS)[15–17] has a principal advantage for the study of localized excitations such as TLSs and LFMs: Since each single molecule is affected *exclusively* by the excitations in its immediate vicinity (of a few nanometers radius), there is no ensemble averaging over the whole sample. This point is essential for the study of strongly inhomogeneous objects such as amorphous solids. For example, the individual parameters of single TLSs[18] and LFMs[19,20] in polymers could be determined from SMS experiments. In the present work we have investigated the low-temperature spectral dynamics of single fluorescent molecules embedded in thin and ultrathin polymer films at low T and the variation of this dynamics as a function of film thickness.

Ultrahin films of polyisobutylene (PIB) weakly doped with fluorescent tetra-*tert*butylterrylene (TBT) molecules were prepared by spin-coating a toluene solution of PIB+TBT onto a microscope coverslip. A very low concentration of TBT was added to the PIB solution. By varying the PIB concentration in the solution we were able to prepare films of varying thickness between 5 and 500 nm. The film thickness and its homogeneity were controlled with a surface profilometer Veeco DEKTAK 150 with 2 nm vertical resolution.

The experiments were carried out with a low-temperature spectro-microscope we built for this purpose (see Refs.[21,22] for more detail). The sample was placed in a He-4 optical cryostat (CryoVac) and cooled down to 4.5 K. The excitation source was a tunable single-mode dye laser Coherent CR 599-21 (operated with Rhodamine 6G) with a linewidth of about 2 MHz and a continuous tuning range of 30 GHz or approximately 0.035 nm. A microscope objective (Mikrothek; NA ~ 0.8) placed inside the cryostat was used to image the red-shifted fluorescence of individual TBT molecules onto a CCD camera (PCO Sensicam QE). Residual excitation light was removed from the detection path with low-pass edge filters.

The fluorescence excitation spectra of a number of individual TBT chromophores were repeatedly measured in parallel by scanning the laser wavelength over a fixed spectral interval of 30 GHz with the camera synchronously recording the single molecule images in the field of view of the microscope objective at a rate of about 10 frames per second. The fluorescence excitation spectrum of a given chromophore molecule is obtained by plotting the integrated intensity of its image versus excitation wavelength. The repeated scanning procedure yields the time-resolved spectra ("spectral histories") of all individual molecules in the field of view simultaneously. Individual spectral trails are usually represented in a two-dimensional plot with the fluorescence instensity (encoded on a gray scale) as a function of excitation wavelength (horizontal axis) and laboratory time (vertical axis)[23] (see Fig. 1).

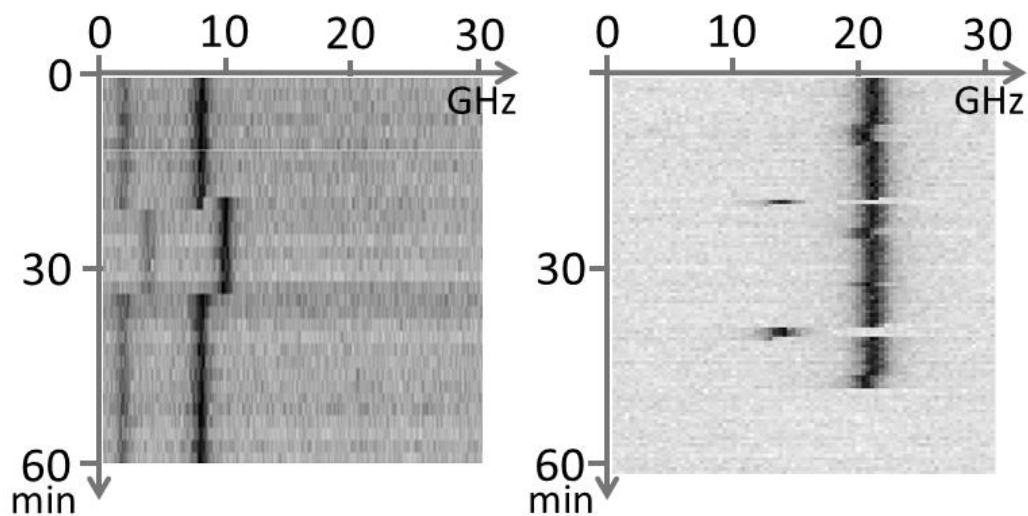

Fig. 1. Two examples of spectral trails of individual TBT molecules in a thick PIB film at 4.5 K. The horizontal axis represents the excitation wavelength, the vertical axis the recording time. The molecule on the right-hand panel shows the coupling to two flipping two-level systems, one causing a larger spectral shift than the other. The molecule on the left panel is also affected by two TLSs, one of them flipping much faster than the recording rate of the spectra.

The individual spectra of fluorescent molecules used for single molecule spectroscopy in solid matrix at cryogenic temperatures exhibit narrow zero-phonon lines (ZPLs). Their narrowness – about $10^{-3}$–$10^{-5}$ nm allows one to detect small spectral shifts such as those caused by the coupling to nearby flipping TLSs in the disordered matrix. There may be different mechanisms causing such shifts. According to the standard tunneling model, a single impurity molecule is expected to switch between $2^N$ spectral positions due to the interaction with $N$ active TLSs which are located near (and strongly coupled to) the molecule. Examples of this "standard" behavior are presented in Fig.1. Smaller spectral shifts due to the "sea" of weakly coupled (e.g., distant) TLSs contribute to the observed spectral linewidth. For a detailed analysis of TLS-impurity interactions at various distances see Ref. [24].

Experiments, however, show that real picture is notably more complex than the model predictions described above. Spectral trails inconsistent with predictions of standard model were observed in small numbers in non-ideal crystals,[25] amorphous polyisobutylene, and polyethylene.[26] In several cases of low-molecular-weight glasses and oligomers this non-standard dynamics strongly prevails.[27,28] Its key features include irreversible jumps, continuous shifting of single-molecule lines, and tunnelling between more than two states (see Fig. 2 and Supplementary Material[29]).

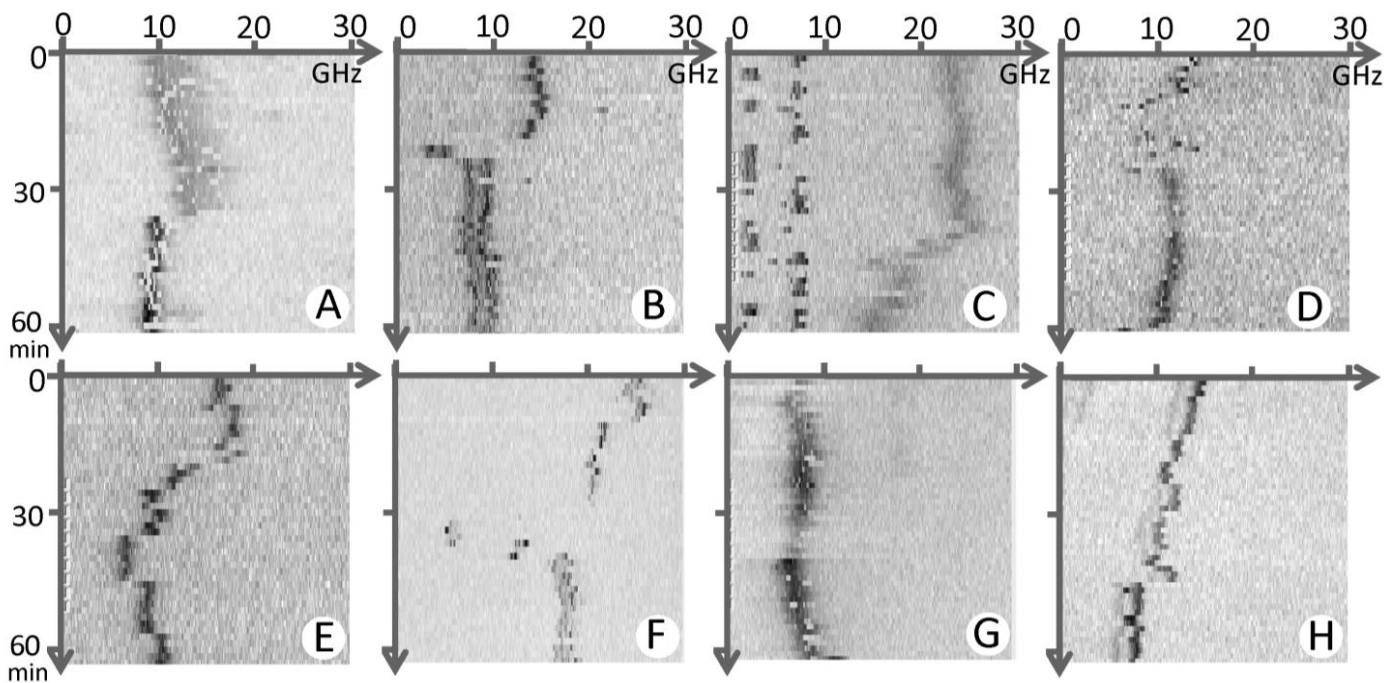

Fig. 2. Selected spectral trails of single TBT molecules in PIB inconsistent with the standard tunneling model. They feature continuous spectral shifts (A, C, E, G, H), abrupt changes of the line width (A, D, G), and random frequency jumps (B, D, F). The temperature was 4.5 K, the excitation wavelength around 584 nm. For more trails see Supplementary Material[29].

In the present study we investigated the spectral dynamics of single TBT molecules in PIB films of different thickness at $T = 4.5$ K. We found that in each film a certain portion of the dopant molecules show spectral dynamics inconsistent with the TLS model and that the relative number of these molecules increases with decreasing film thickness. Also the excitation intensity of the laser has an influence, with higher intensities enhancing the portion of molecules with non-standard behavior, especially in the thinnest films (see Fig.3). On the other hand, in PIB films of all thicknesses also a certain number of molecules are present whose spectral dynamics is in agreement with the standard tunnelling model.

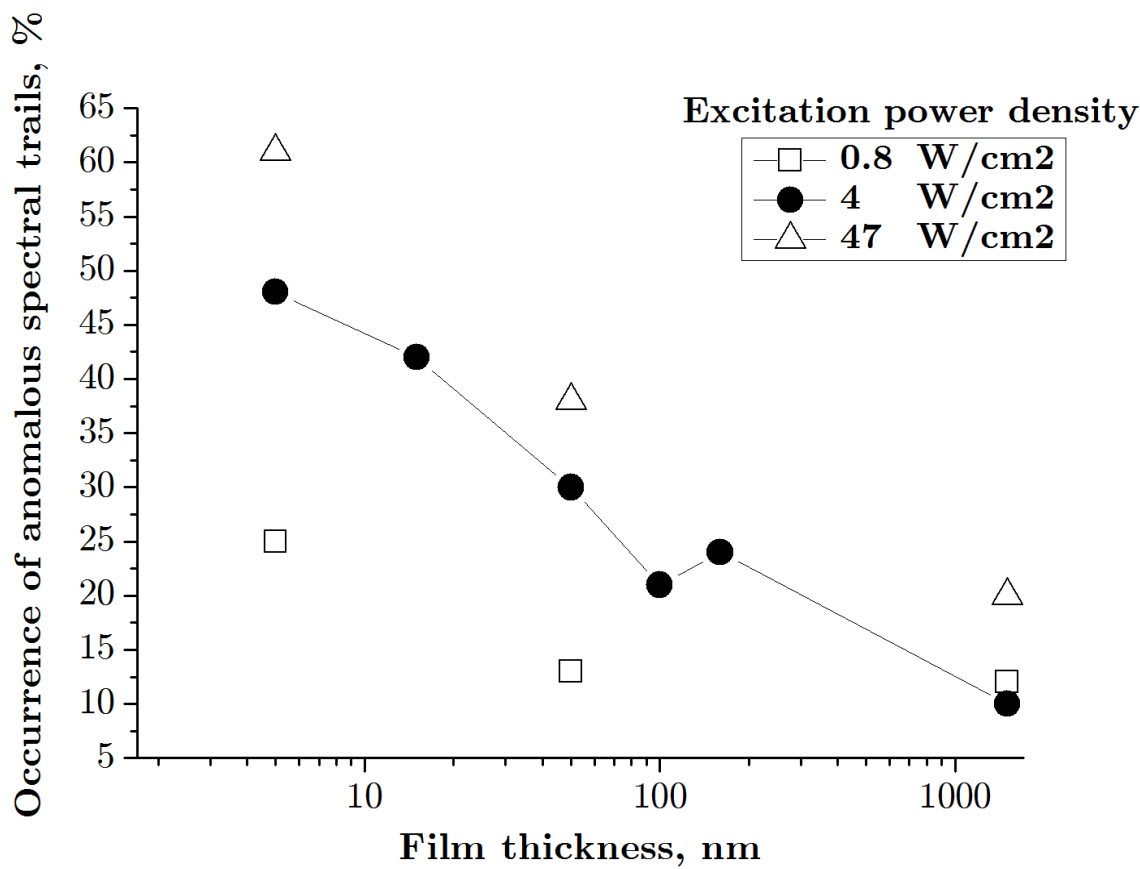

Fig.3. Relative number of single TBT molecules in thin PIB films exhibiting non-standard spectral dynamics as a function of film thickness. The experiments were performed with three different laser intensities. Standard errors are not shown as they are not well defined in this case (see text for details).

All "non-standard" spectral trails can be roughly divided into three classes showing "different degrees of incompatibility" with the standard tunnelling model: The first two groups require modified distributions of TLS parameters for PIB, whereas the third class indicates the presence of interactions between TLSs. In the first group, the spectral line visits the scanned 30 GHz interval only briefly – this implies that there may be unusually numerous tunneling systems in the vicinity of the fluorescent molecule whose transitions induce large spectral shifts to its line. The second class is characterized by continuous drifts of the SM line, mainly in one direction. This may also be explained by a large number of active, yet more distant, TLSs systems around the molecule, causing many small independent spectral shifts. In fact, some of the first spectral trails ever recorded in single-molecule spectroscopy (pentacene in crystalline *p*-terphenyl) featured a similar type of spectral-drift behavior,[25] yet, they were shown to be in excellent agreement with the tunneling model, if a special type of tunneling system is assumed to cause the spectral dynamics.[30] Finally, in the third group of trajectories the SM lines are subject to sudden changes of the linewidth (Fig. 2A, D, G) or the jump rate (Fig. 2B, D, F). It seems that this behavior cannot be explained by a model of independent TLSs; interactions between them must be taken into account.

It is important to note that when concerning trails from first and second groups (as described in previous paragraph), in some cases it is unclear whether the trail in question really belongs to these first two "non-standard" groups or can be fit into the standard tunneling model as it is. Occurrence of these ambiguous trails – about 15% – is the main source of error in Fig. 3, where all recorded trails are viewed as either trails that fit into the standard TLS model or those that do not. Our attempts to construct a robust statistical criterion for classification of trails yielded results that were far less stable than results of manual sorting.

Continuous spectral drifts of single-molecule lines can also be attributed to slow structural relaxations, as noted by Boiron et al.[26] Structural relaxations were proven to exist in polymers at low temperatures,[31] they are thought to be, in general, of different nature than those occurring near the glass transition – because at low temperature thermal energy is lower than most energy barriers, while at glass transition temperature these two values are comparable. In ultrathin films, however, additional structural relaxations may be related to low energy barriers due to influence of free surface – similar to the barrier-lowering influence of free surfaces observed near the glass transition temperature. Another possible way to model low-temperature structural relaxations is the cascades of tunnelling events in interacting two-level systems.

Increased mobility near the polymer surface may be caused by loops of polymer chains that reach the free surface and are be able to transport voids of free volume from the surface into the interior of the film.[32] Although studies of the glass transition temperature in thin films do not confirm this mechanism,[33] its influence may be larger at cryogenic temperatures.

The hypothetical higher spatial density of "active" tunneling systems or the occurrence of special types of structural relaxations in ultrathin films would also present the possibility of explaining those spectral trails which feature abrupt changes of their spectral parameters such as linewidth and jump rate (Fig. 2). These features cannot be produced by any collection of *independent* TLSs, but they can be explained by a combination of few *interacting* tunneling systems. One possibility would be, e.g., that flips of one TLS alter the barrier height of a neighboring TLS and, in this way, dramatically affect the jumping rate of the latter. A high spatial density of active tunneling systems is expected to lead to an increased number of interacting TLS pairs. Alternatively, also structural relaxations may be able to affect TLS parameters.

Experiments have also shown that the relative number of non-standard spectral trails increases under stronger laser excitation (se Fig. 3). We estimate that the absorbed power density is still not sufficient to heat the sample significantly, especially close to the surface. Thus, the larger contribution of light-induced dynamics in ultrathin films probably arises from non-equilibrium processes, e.g. phonons emitted by the fluorescent molecule travel a few nanometers and cause additional flips of nearby tunneling systems with low barriers before being thermalized. Moreover, the proximity of the free surface in ultrathin films may provide conditions – such as surface modes – that extend the traveling range of non-equilibrium phonons.

In conclusion, the spectral low-temperature dynamics of single chromophore molecules embedded in films of amorphous polyisobutylene with thicknesses ranging from 300 nm down to 5 nm was investigated. We observed a gradual increase of the fraction of single molecules exhibiting non-standard spectral dynamics with decreasing thickness of the polymer film. The effect is also sensitive to the intensity of the excitation light. We tentatively ascribed our observations to a higher density and/or modified parameter distributions of tunneling systems close to the free surface. Also the occurrence of structural relaxations and the coupling of non-equilibrium phonons to surface modes may play a role.

The authors would like to thank A. Krekhov and I. Eremchev for fruitful discussions and the members of the chair Experimental Physics IV in Bayreuth for invaluable experimental support. This work was funded by the Deutsche Forschungsgemeinschaft, the Russian Foundation for Basic Research (13-02-00919, 11-02-00816, 12-02-33027), and a Grant of the President of Russia (MD-465.2012.2).